\documentclass[twocolumn,
showpacs,showkeys,nofootinbib,amsfonts,amssymb
,floatfix
]{revtex4-1}

\usepackage{graphicx}  
\usepackage{amsmath, amssymb, bm,bbm}   % AMS packages  
\usepackage{natbib}
\usepackage{hyperref}

\usepackage{longtable}
\usepackage{csvsimple} %Load CSV file as tables
\usepackage{adjustbox} %  Fit wide table

\usepackage{mathtools}
\usepackage{ifthen}

\def\){\right)} 
\def\({\left(} 
\def\]{\right]} 
\def\[{\left[}

\def\TrAlpha{$^3\mathrm{H}(\alpha,\gamma)^7\mathrm{Li}$}
\def\HeAlpha{$^3\mathrm{He}(\alpha,\gamma)^7\mathrm{Be}$}

\def\HeAlphaHead{$\bm{^3}$H\lowercase{e}$\bm{(\alpha,\gamma)^7}$B\lowercase{e}}
\def\TrAlphaHead{$\bm{^3}$H$\bm{(\alpha,\gamma)^7}$L\lowercase{i}}

\begin{document}

\title{Bayesian analysis of capture reactions \texorpdfstring{\HeAlphaHead}{He3-Alpha} and \texorpdfstring{\TrAlphaHead}{H3-Alpha}}

\author{%
Pradeepa Premarathna}
\email{psp63@msstate.edu}

\author{%
Gautam Rupak}
\email{grupak@ccs.msstate.edu}

\address{Department of Physics \& Astronomy and HPC$^2$ Center for 
Computational Sciences, Mississippi State
University, Mississippi State, MS 39762, USA}

\begin{abstract}
Bayesian analysis of the radiative capture reactions  \HeAlpha ~and \TrAlpha ~are performed to draw inferences about the cross sections at threshold. We do a model comparison of two competing effective field theory power countings for the capture reactions. The two power countings differ in the contribution of two-body electromagnetic currents. In one power counting, two-body currents contribute at leading order, and in the other they contribute at higher orders. The former is favored for \HeAlpha ~if elastic scattering data in the incoming channel is considered in the analysis. Without constraints from elastic scattering data, both the power countings are equally favored.  For \TrAlpha, the first power counting with two-body current contributions at leading order is favored with or without constraints from elastic scattering data. 
 \end{abstract}

%\date{\today}
\maketitle

%===================================================
\section{Introduction}
\label{sec:intro}
%===================================================
Effective field theory (EFT) calculations of low-energy capture reactions provide important uncertainty estimates in cross section evaluations.  Element synthesis in the Big Bang and in stellar environments involve reactions at low energies. These reaction rates are suppressed by the Coulomb repulsion at low energy, making their measurements challenging. For example, the proton-proton fusion rate is not known experimentally at the relevant solar energy~\cite{Adelberger:2010qa}. Theory is needed to extrapolate measurements at higher energies to the astrophysically relevant lower energies. EFT extrapolations are attractive as they provide model-independent error estimates.  The basic idea in the EFT formulation is that one only considers certain degrees of freedom as dynamical at low momentum $p$ below a cutoff $\Lambda\gg p$.  The physics associated with the high-momentum scale $\Lambda$ is not modeled but contributes through dimensionful couplings in the low-momentum theory. Observables are calculated as an expansion in $Q/\Lambda\ll 1$ called the power counting. $Q\sim p$ characterizes the low-momentum physical scale. Theoretical uncertainties are estimated from powers of $Q/\Lambda$ at a given order of the perturbative expansion.

Typical radiative capture reactions such as   \HeAlpha ~and \TrAlpha ~considered in this work depend on initial state interaction, electro-weak currents and the final state wave function. Information about the initial state interaction at low momentum can be incorporated by matching the EFT couplings to elastic scattering phase shifts. The final state physics is contained in the bound state energy and the asymptotic wave function normalization which can be determined from the relevant phase shift as well. The one-body electro-weak currents are well determined, and usually they give the dominant contribution.  A difficulty in these calculations is that the low momentum phase shifts are poorly known which introduces large uncertainty in the construction of the EFT interactions. Moreover, in systems that involve weakly-bound states the couplings can be fine tuned such that two-body currents contribute at leading order (LO) of the perturbation~\cite{Higa:2016igc}.  

The  low-energy capture reactions \HeAlpha ~and \TrAlpha ~are of great interest in nuclear astrophysics. The reactions  \HeAlpha$(p,\gamma)^8$Be ~provide the high energy $^8$Be neutrinos detected at Super-K~\cite{Fukuda:2001nj} and SNO~\cite{Aharmim:2006kv}. The neutrino flux is proportional to the \HeAlpha ~cross section $\sigma$ near its Gamow energy $E_G\sim 20$ keV as $\sigma^{0.81}$~\cite{Bahcall:1987jc,Cyburt:2008up}. Several measurements of the   \HeAlpha ~capture rate have been performed in recent decades, for example ~\cite{ERNA,LUNA,NotreDame,Seattle,Weizmann}. The reaction 
\TrAlpha ~is important for calculating $^7$Li abundances, constraining astrophysical models of Big Bang Nucleosynthesis~\cite{kawano:1991ApJ372,Smith:1993ApJS85219S}. There are less constraints on this reaction from recent measurements~\cite{Brune:1994,Tomsk:2017}.

In this work we use Bayesian analysis to revisit the calculation of  \HeAlpha~\cite{Higa:2016igc}, and also extend the analysis to the related reaction \TrAlpha.  Bayesian analysis is useful in situations as these where we have a large number of parameters that are poorly constrained by lack of accurate data especially in the elastic channels. Parameter estimation is important in not only extrapolation of reaction rates to solar energies but also to develop a systematic power counting expansion in $Q/\Lambda$. This is crucial for providing theoretical error estimates. Further, Bayesian analysis allows for a quantitative comparison of EFT formulations where, for example, the two-body currents appear at different orders in the perturbation. 

The expressions for the cross sections are from Ref.~\cite{Higa:2016igc} that used halo EFT in the calculations. In halo EFT~\cite{Bertulani:2002sz,Bedaque:2003wa}, nuclear clusters are treated as point like particles at momenta that are too small to cause the clusters to break apart. Capture reaction calculations are outlined in the review article Ref.~\cite{Rupak:2016mmz}. We consider two power countings depending on the size of the scattering length $a_0$ in the incoming $s$-wave channel.  In one power counting, two-body currents contribute at LO and in the other they contribute at next-to-leading order (NLO). 

%===================================================
\section{Halo EFT calculation}
\label{sec:EFT}
%===================================================
The calculations for  \HeAlpha ~and \TrAlpha ~are similar.  The discussion here follows 
Ref.~\cite{Higa:2016igc}. The ground state of $^7$Be has a  binding energy $B_0^\mathrm{(Be)}=1.5866$ MeV and spin-parity assignment $\frac{3}{2}^-$. The first excited state has a  binding energy $B_1^\mathrm{(Be)}=1.1575$ MeV and spin-parity assignment $\frac{1}{2}^-$. These binding energies are identified with the low-energy scale~\cite{Higa:2016igc} and they are treated as a bound state of $^3$He and $\alpha$ clusters in the calculation.  The higher excited states of $^7$Be with the same spin-parity as the first two states lie at least   about 8 MeV above the $^3$He-$\alpha$ threshold. These are not included in the low-energy theory. 
The proton separation energy of $^3$He ($\sim 5.5$ MeV), excited states of $\alpha$ ($\gtrsim20$ MeV) are identified with the high-energy scale as well. Thus, we treat the $^3$He and $\alpha$ clusters as point-like fundamental particles. The spin-parity assignments of $\frac{1}{2}^+$ for the $^3$He and $0^+$ for the $\alpha$ implies that the $\frac{3}{2}^-$,
$\frac{1}{2}^-$  states of $^7$Be are $p$-wave bound states of $^3$He and $\alpha$. Then radiative capture at low energy is dominated by E1 transition from $s$-wave initial states with a small contribution from the $d$ waves~\cite{Higa:2016igc}. For the  \TrAlpha ~reaction, the ground state of $^7$Li has a binding energy 
$B_0^\mathrm{(Li)}=2.467$ MeV and spin-party $\frac{3}{2}^-$. The first excited state has a  binding energy $B_1^\mathrm{(Li)}=1.989$ MeV and spin-parity assignment $\frac{1}{2}^-$. These two $p$-wave states of the 
$\frac{1}{2}^+$ $^3$H and $0^+$  $\alpha$  are included in the theory. The higher excited states with the same quantum numbers lie at least about 6 MeV above the $^3$H-$\alpha$ threshold. These are not included in the low-energy theory. The neutron separation energy of $^3$H ($\sim 6$ MeV) is identified with the high energy theory, and so $^3$H is treated as a point-like fundamental particle along with the $\alpha$.  The main differences between the EFT expressions for the \HeAlpha ~and \TrAlpha ~cross sections are the charges of the nuclei, and the relevant masses and binding energies. The separation between the low and high energy is smaller in the $^3$H-$\alpha$ system which would result in a slower  convergence of the  $Q/\Lambda$ expansion.

The total cross section for both the capture reactions is given as~\cite{Higa:2016igc}:
 %-----------  Equation 1 --------------
\begin{multline}\label{eq:sigma}
  \sigma(p)=\frac{1}{16\pi M^2}\frac{1}{2}
  \left[ \frac{p^2+\gamma_1^2}{2\mu p} |M^{(^2P_{1/2})}|^2 \right.\\
  \left. + \frac{p^2+\gamma_0^2}{2\mu p}|M^{(^2P_{3/2})}|^2  \right],
\end{multline}
at center-of-mass (cm) relative momentum $p$ of the incoming particles. 
We use the spectroscopic notation $^{2s+1}L_{j}$ to indicate the capture to the $^2P_{3/2}$ ground state with binding momentum $\gamma_0=\sqrt{2\mu B_0}$, and  the $^2P_{1/2}$ first excited state with binding momentum $\gamma_1=\sqrt{2\mu B_1}$. $M= M_\psi+M_\phi$ is the total mass, and $\mu=M_\psi M_\phi/M$ is the reduced mass of the incoming particles. $M_\phi$ is the scalar $0^+$ $\alpha$ mass, and $M_\psi$ is the fermion $\frac{1}{2}^{+}$ $^3$He or $^3$H mass as appropriate. The squared amplitude is given by
%-----------  Equation 2 --------------
\begin{multline}
  |M^{(\zeta)}|^2=(2j+1)
  \left(\frac{ Z_\phi M_\psi}{M}-\frac{ Z_\psi M_\phi}{M}\right)^2\\
\times  \left[ |\mathcal A(p)|^2 +2|Y(p)|^2\right]  \frac{64\pi\alpha_e  M^2 ([h^{(\zeta)}]^2 \mathcal Z^{(\zeta)})}{\mu}\, ,
\end{multline}
where $\zeta= {}^2P_{3/2}$ for the ground state ($j=3/2$) and $\zeta={}^2P_{1/2}$ for the exited state ($j=1/2$). $Z_\phi=2$ is the charge of the $\alpha$ and $Z_\psi$ is the charge of either the $^3$He or the $^3$H  particle. $\alpha_e=e^2/(4\pi)\approx 1/137$ is the fine structure constant. 
 The capture from initial $s$ and $d$ wave states are given by the amplitudes $\mathcal A(p)$ and $Y(p)$ shown in the Appendix, Eq.~(\ref{eq:swaveA}) and Eq.~(\ref{eq:dwaveA}) respectively.  The factor $([h^{(\zeta)}]^2 \mathcal Z^{(\zeta)})$ is associated with the normalization of the outgoing $^7$Be or $^7$Li particle wave function, shown in Appendix, Eq.~(\ref{eq:Zphi}). They are related to the asymptotic normalization constant (ANC) as:
 %%----------- Equation 3 --------------
\begin{align}
C_{1,\ \zeta}^2=\frac{[\gamma^{(\zeta)}]^2 \Gamma(2+k_C/\gamma^{(\zeta)})}{\pi} [h^{(\zeta)}]^2 \mathcal Z^{(\zeta)}.
\end{align}
 The ANCs depend on the binding momentum and the effective range $\rho_1$ of the corresponding $p$-wave elastic channel of the outgoing $^7$Be or $^7$Li state.

 The $S$-factor is calculated as a function of the c.m. incoming kinetic energy
 $E=p^2/(2\mu)$ from the cross section as $S_{34}(E)=E e^{2\pi\eta_p}\sigma(p=\sqrt{2\mu E})$ where $\eta_p=\alpha_e Z_\psi Z_\phi\mu/p$ is the Sommerfeld parameter.  

%===================================================
\section{Bayesian parameter estimation and model comparison}
\label{sec:Bayes}

In the EFT calculation we have several unknown couplings and parameters that we wish to constrain from available data. 
Suppose we represent the set of unknown couplings and parameters by the vector $\bm\theta$ in the multi-dimensional parameter space. In the frequentist approach, the parameters $\bm\theta$ are determined by minimizing the $\chi^2$ given by
%----------- Equation 4 --------------
\begin{align}
\chi^2 &= \sum_{i=1}^N\frac{[y_i-\mu_i(\bm\theta)]^2}{\sigma_i^2}\, ,
\end{align}  
where the data set $D$ consists of the $N$ measurements $y_i$ with corresponding measurement errors $\sigma_i$. $\mu_i(\bm\theta)$ are the theory predictions. Minimizing $\chi^2$ maximizes the likelihood function  $P(D|\bm\theta,H)\propto\exp(-\chi^2/2)$ which gives the conditional probability of the data set $D$ to be true given the parameters $\bm\theta$ and the proposition $H$. Background information, theoretical assumptions, and hypothesis are included in the proposition $H$. 

In Bayesian analysis, the question one wants to answer is how likely some parameter value $\bm\theta$ given the data $D$. The answer is contained in the posterior probability distribution given by
%----------- Equation 5 --------------
\begin{align}\label{eq:posterior}
P(\bm\theta|D,H)= \frac{P(D|\bm\theta,H)P(\bm\theta|H)}{P(D|H)}\, ,
\end{align} 
where $P(\bm\theta|H)$ is the prior distribution of the parameters which is predicated on the theoretical assumptions that is made in proposition $H$. The prior distribution is where one can, for example, include EFT assumptions about the expected sizes of the unknown couplings and parameters. The evidence $P(D|H)$ is also known as the marginal likelihood because
%----------- Equation 6 --------------
\begin{align}\label{eq:evidence}
P(D|H)= \int \prod_{i}d\theta_i  P(D|\bm\theta,H)P(\bm\theta|H),
\end{align}
comes from marginalizing the likelihood on the parameters.  We used $\int \prod_i d\theta_i P(\bm\theta|D,H) =1$ above for normalized probabilities.

Parameter estimation from the posterior distribution in Eq.~(\ref{eq:posterior}) does not require evaluation of the overall normalization factor, the evidence $P(D|H)$. Markov chain Monte Carlo (MCMC) algorithms such as Metropolis-Hastings or some variation of it can be used to sample points from the posterior distribution $P(D|\bm\theta,H)P(\bm\theta|H)$. The MCMC algorithm takes a prior $P(\bm\theta|H)$ based on some physics principle, and combines it with the data where the likelihood $P(D|\bm\theta,H)$ is maximized to produce the posterior distribution. Each sample in the MCMC simulation represents  a point in the multi-dimensional parameter space. The calculation of the  marginalized probability distribution of a parameter say $\theta_1$ 
%----------- Equation 7 --------------
\begin{align}
P(\theta_1|D,H)=\int \prod_{i\neq 1} d\theta_i P(\bm\theta|D,H)\, ,    
\end{align}
is straightforward: just look at the distribution of the coordinate $\theta_1$ in the sample which trivially corresponds to summing over all the other coordinates~\cite{Brewer:2018}.

Model comparison in Bayesian analysis requires calculating the relevant evidences. In $\chi^2$ minimization, one might get a better fit by including extra parameters. However, in Bayesian analysis there is a cost associated with including extra parameter as it introduces another prior distribution. The probability ratio of model A ($M_A$) and 
model B ($M_B$), for a data set $D$, is given by the posterior odd ratio:
%----------- Equation 8 --------------
\begin{align}\label{eq:postodd}
\frac{P(M_A|D,H)}{P(M_B|D,H)}=\frac{P(D|M_A,H)}{P(D|M_B,H)}\times \frac{P(M_A|H)}{P(M_B|H)}\, .
\end{align}
The first factor on the right is just the ratio of evidences for $M_A$ and $M_B$. The second factor is called the prior odd.  In our analysis, when we compare models we assume no prior bias towards or against any model, and accordingly set the prior odd to 1.

The evidence calculation requires numerical integration in the multi-dimensional parameter space. If we label the parameters of $M_A$ by the vector $\bm\alpha$, and the parameters of $M_B$ by the vector $\bm\beta$, then the 
evidences are:
%----------- Equation 9 --------------
\begin{align}
P(D|M_A,H)  &=   \int \prod_i d \alpha_i  P(D|\bm\alpha,M_A,H)P(\bm\alpha| M_A,H)\, , \nonumber\\
P(D|M_B,H) &=   \int \prod_i d \beta_i  P(D|\bm\beta,M_B,H)P(\bm\beta| M_B,H)\, .
\end{align}

The difficult task of calculating the evidences in the multi-dimensional parameter space is made easier in the Nested Sampling method introduced by Skilling~\cite{Skilling:2006}.
The basic idea is to draw a set of points randomly from the prior
distribution $P(\bm\theta, H)$, and arrange them from best to worst in terms of the likelihood function. Then one discards the worst point, and replaces it with another point from the prior distribution with the condition that it has a larger likelihood value. This guarantees that progressively higher likelihood regions of the parameter space is probed. 
In the original work by Skilling, at every iteration the worst point was replaced by Monte Carlo update using the Metropolis-Hastings algorithm. There
has been a lot of algorithmic development since then on how to update the list of
points, see review article~\cite{Brewer:2018}. We tried several of these methods. For our expressions, these methods disagree by about $\sim 2$ in the calculation of $\ln[P(M_A|D,H)/P(M_B|D,H)]$, and we assume this level of systematic errors in model comparisons. We decided to use the MultiNest  algorithm~\cite{Feroz:2009} implemented in Python~\cite{Nestle}. We cross checked the numbers with our own Fortran code using the original Nested Sampling algorithm~\cite{Skilling:2006}.

%===================================================
\section{\texorpdfstring{$\bm{S}$-factor}{S-factor} estimation for \texorpdfstring{\HeAlphaHead}{He3-Alpha} and \texorpdfstring{\TrAlphaHead}{H3-Alpha}}
\label{sec:Sfactor}

The cross section (and the related $S$-factor) expressions for \HeAlpha ~and \TrAlpha 
~are given by  Eq.~(\ref{eq:sigma}). It depends on the amplitudes $\mathcal A(p)$ and $Y(p)$. The latter is expressed in Eq.~(\ref{eq:dwaveA}) without any unknown parameters. The former depends on the initial state strong interaction $s$-wave phase shift $\delta_0$ that can be written  model-independently in terms of scattering parameters: scattering length $a_0$, effective range $r_0$, shape parameter $s_0$, etc., in 
Eq~(\ref{eq:swaveERE}). We keep only three $s$-wave scattering parameters $a_0$, $r_0$ and $s_0$ in our analysis. $\mathcal A(p)$ also depends on 2 two-body current couplings  $L^{(\zeta)}_\mathrm{E1}$ for $\zeta={}^2P_{3/2},\ \zeta={}^2P_{1/2}$. The cross section is multiplied by the overall factor $([h^{(\zeta)}]^2 \mathcal Z^{(\zeta)})$, Eq.~(\ref{eq:Zphi}), that depends on the binding momenta $\gamma$ and the effective ranges $\rho_1$ in the $p$-wave channels $^2P_{3/2}$ and $^2P_{1/2}$. This amounts to 7 parameters/couplings for the capture cross section as we take the binding momenta as given. We use the notation $\pm$ to indicate the  ${}^2P_{3/2}$ and ${}^2P_{1/2}$ channels, respectively. 

A $\chi^2$ fit to capture data~\cite{Higa:2016igc} for \HeAlpha ~does not constraint the $p$-wave effective ranges $\rho_1^{(\pm)}$ accurately. The wave function normalization constants $[h^{(\pm)}]^2 \mathcal Z^{(\pm)}$ have poles at approximately $\rho_1^{(+)}=-47.4$ MeV and $\rho_1^{(-)}=-32.4$ MeV, respectively. A small change in the effective range values changes the normalization of the cross section significantly. Moreover, one finds that for the parameters from the $\chi^2$ fit, there is a large cancellation among the terms in $\mathcal A(p)$ such that the two-body contribution becomes important. However, the contribution from the two-body currents can be suppressed by choosing the effective range contribution near the pole in $([h^{(\zeta)}]^2 \mathcal Z^{(\zeta)})$.  Therefore, we find it important to include information from $p$-wave elastic phase shifts $\delta_1^{(\pm)}$ to constraint $\rho_1^{(\pm)}$ more accurately. This entails including the shape parameters $\sigma_1^{(\pm)}$ in the fit. $\sigma_1^{(\pm)}$ contributes to phase shifts $\delta_1^{(\pm)}$ but not to the capture cross section. The total parameters to be fit now increases to 9: $a_0$, $r_0$, $s_0$, $\rho_1^{(\pm)}$, $\sigma_1^{(\pm)}$, $L_\mathrm{E1}^{(\pm)}$. 

We start with the new analysis of \HeAlpha~\cite{Higa:2016igc}. The capture and elastic scattering data is more accurately known for this reaction than for \TrAlpha.

%===================================================
\subsection{Capture reaction \texorpdfstring{\HeAlphaHead}{He3-Alpha}}

We use charge $Z_\psi=2$, mass $M_\psi=2809.41$ MeV for the incoming $^3$He, and 
charge $Z_\phi=2$, mass $M_\phi=3728.4$ MeV for the incoming $\alpha$ in Eq.~(\ref{eq:sigma}). 
The \HeAlpha ~capture data are from ERNA~\cite{ERNA}, LUNA~\cite{LUNA}, Notre Dame~\cite{NotreDame}, Seattle~\cite{Seattle}, and Weizmann~\cite{Weizmann}. 
It includes prompt photon, activation and recoil (ERNA) measurements. 
For the prompt data, the branching ratio $R_0$ of capture to the excited state to 
the ground state is available. 

In the Bayesian analysis we explore fits to capture data up to c.m. energy (momentum)  $E\lesssim 1$ MeV ($p\lesssim 60$ MeV) that we call region I. Fits to capture data over  $E\lesssim 2$ MeV ($p\lesssim 80$ MeV) we call region II. 
The phase shift data for $\delta_0$, $\delta_1^{(\pm)}$ 
is from an old source~\cite{Spiger:1967} that was 
analyzed by Boykin {\em et al.} in Ref.~\cite{Boykin:1972}. The phase shift data starts from around $E\sim 2$ MeV ($p\sim 80$ MeV), which is at the higher end of the range of applicability of the EFT. We fit with phase shift data, and also without (indicated by an asterisk $\ast$ as appropriate). In all, we have four combinations of data sets for the fits.  Region I includes 42 $S$-factor data and 20 $R_0$ data points whereas region II includes 70 $S$-factor data and 32 $R_0$ data points. The phase shifts include 10 data points in each of the three channels.  

In the EFT, the momentum $p\sim 70$ MeV and final state binding momenta $\gamma_0\sim\gamma_1\sim 60\mbox{--}70$ MeV constitutes the small scale $Q$. The pion mass, excited states of $^3$He, $\alpha$, etc., constitutes the cutoff scale $\Lambda\gtrsim 150\mbox{--}200$ MeV. In the $s$-wave capture amplitude $\mathcal A(p)$, 
Eq.~(\ref{eq:swaveA}), the first term $X(p)$ that only contains Coulomb interaction is $O(1)$ at small momentum. The rest of the contributions, at arbitrarily small momentum $p$, scales as $a_0 (B+\mu J_0)/\mu^2$ and  $a_0 k_0 L_\mathrm{E1}^{(\zeta)}$. The linear combination $(B+\mu J_0)/\mu^2$ scales as $Q^3/\Lambda^2$, and the photon energy $k_0=(p^2+\gamma^2)/(2\mu)\sim Q^3/\Lambda^2$~\cite{Higa:2016igc,Higa:2008dn}. Thus at low momentum the contributions from initial state strong interaction 
$a_0 (B+\mu J_0)/\mu^2\sim a_0 Q^3/\Lambda^2$. The two-body current contribution 
$a_0 k_0 L_\mathrm{E1}^{(\zeta)}\sim a_0 Q^3/\Lambda^2$ for a natural sized dimensionless coupling $L_\mathrm{E1}^{(\zeta)}\sim 1$. The size of the $s$-wave scattering length then determines the relative contributions in the amplitude at low momentum.

%%%------------- Figure  1 --------------------------------
\begin{figure}[tbh]
\begin{center}
\includegraphics[width=0.47\textwidth,clip=true]{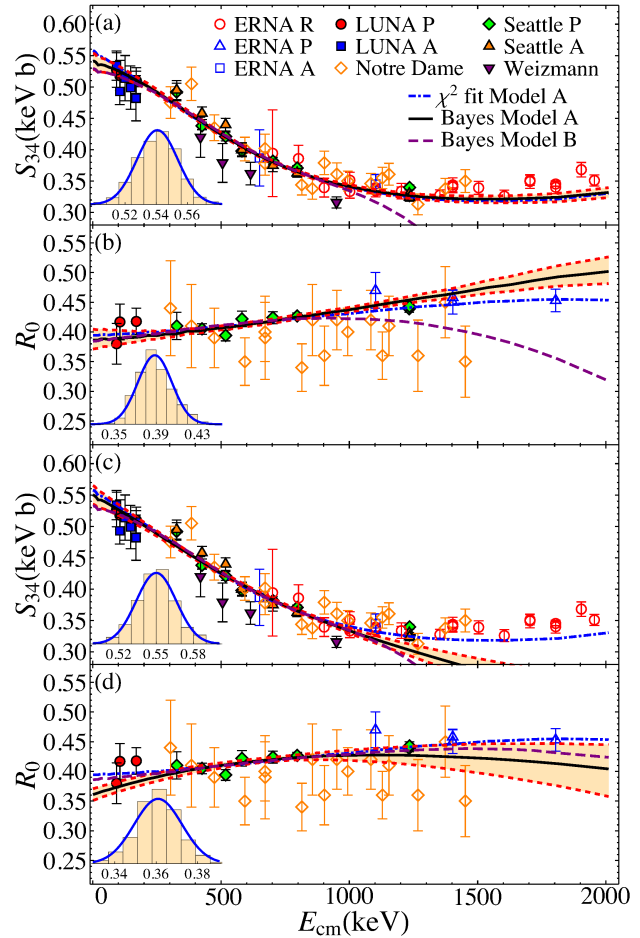}
\end{center}
\caption{\protect \HeAlpha: Model comparison for capture cross section. Solid (black) curve is the median result using Model A. The shaded region, bounded by the dashed (red) curves, represents $68\%$ of the posterior distribution. The long-dashed (purple) curve is for Model B. We include the $\chi^2$ result as dot-dashed (blue) curve for comparison. Panel (a) and (b) are simultaneous fits to capture and phase shift data. Panel (c) and (d) are fits only to capture data. These fits are to capture data in region I, $E\lesssim 1000$ keV. The inset shows the distribution for Model A at $E_\star=60\times 10^{-3}$ keV.}
\label{fig:Be7capture}
\end{figure}

$\chi^2$ minimization that included two-body current contribution gave a large $a_0\sim 20\mbox{--}30\ \mathrm{fm}\sim \Lambda^2/Q^3$, and consequently the initial state interactions 
$a_0 (B+\mu J_0)/\mu^2\sim 1$ and two-body currents $a_0 k_0 L_\mathrm{E1}^{(\zeta)}\sim 1$ are LO contributions. $\chi^2$ minimization without two-body currents give a smaller $a_0\sim 5-10\ \mathrm{fm}\sim\Lambda/ Q^2$~\cite{Higa:2016igc}. This will make initial state interaction  $a_0 (B+\mu J_0)/\mu^2\sim Q/\Lambda$ a NLO contribution. In EFT, there is always going to be two-body currents unless some symmetry prevents them. Then the only consideration is at what order in perturbation they contribute. Again assuming natural sized couplings $L_\mathrm{E1}^{(\zeta)}\sim 1$, the two-body contribution $a_0 k_0 L_\mathrm{E1}^{(\zeta)}\sim Q/\Lambda$ is also a NLO effect for $a_0\sim \Lambda/Q^2$.

%%%------------- Figure  2 --------------------------------
\begin{figure}[tbh]
\begin{center}
\includegraphics[width=0.47\textwidth,clip=true]{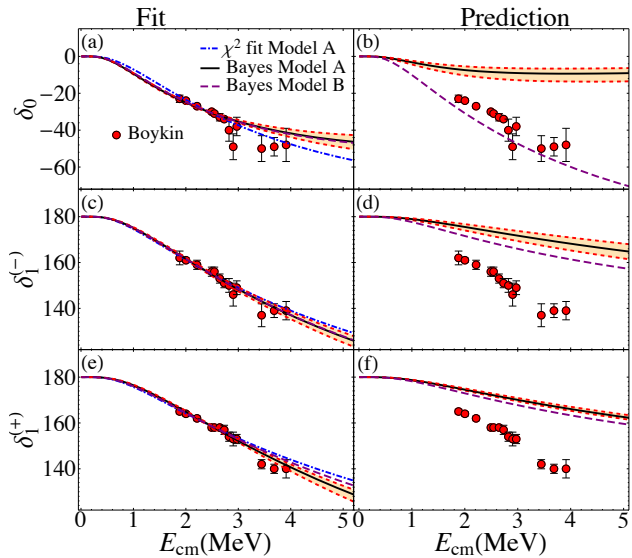}
\end{center}
\caption{\protect\HeAlpha: Model comparison for  scattering phase shifts. Notation for the curves 
the same as in Fig.~\ref{fig:Be7capture}. Panels on the left are fits to phase shifts, and the panels on the right are predictions as discussed in the text.}
\label{fig:Be7PhaseShift}
\end{figure}

We consider two power countings. The first one assumes a large $a_0\sim \Lambda^2/Q^3$. This we call ``Model A" though EFT is a model-independent formulation. In this power counting~\cite{Higa:2016igc}, all the other parameters and couplings are natural sized determined by their naive dimensions $r_0\sim 1/\Lambda$, $s_0\sim 1/\Lambda^3$, $\rho_1^{(\pm)}\sim \Lambda$, $\sigma_1^{(\pm)}\sim 1/\Lambda$, $L_\mathrm{E1}^{(\pm)}\sim 1$. The capture cross section depends on $a_0$, $r_0$, $\rho_1^{(\pm)}$, $L_\mathrm{E1}^{(\pm)}$ at LO. The NLO contribution gets an additional contribution from $s_0$. The $s$-wave phase shift $\delta_0$ depends on $a_0$ and $r_0$ at LO, and at NLO brings in $s_0$. 
A fine-tuning in the linear combination $r_0 p^2/2-2 k_C H(\eta_p)\sim Q^3/\Lambda^2$ promotes the effective range contribution to LO~\cite{Higa:2016igc}.  
The $p$-wave phase shifts $\delta_1^{(\pm)}$ depends on $\rho_1^{(\pm)}$ at LO and the NLO contribution comes from $\sigma_1^{(\pm)}$. The $d$ wave contribution $Y(p)$ is included at NLO due to a large near cancellation at LO~\cite{Higa:2016igc}. 

In the second power-counting that we call ``Model B", one assumes a smaller $a_0\sim \Lambda/Q^2$. The rest of the parameters have the same scaling as Model A. However, the perturbative expansion is now different. The LO capture cross section has no initial state strong interaction. It only depends on $\rho_1^{(\pm)}$. $a_0$ and $L_\mathrm{E1}^{(\pm)}$ contributes at NLO, and $r_0$ at next-to-next-to-leading order (NNLO). In phase shift $\delta_0$, 
$a_0$ contributes at LO, $r_0$ at NLO and $s_0$ at NNLO. The $p$-wave phase shift contributions remain unchanged from Model A above. The $d$ wave contribution $Y(p)$ is now included at NNLO.

Model A and Model B are compared by calculating the posterior odd ratio $P(M_A|D,H)/P(M_B|D,H)$ from Eq.~(\ref{eq:postodd}). The $\chi^2$ for the fit without  two-body currents was larger than the one with two-body currents when phase shift data, especially $\delta_0$, was used. 
We explore the possibility that the uncertainty in the phase shifts are actually larger than estimated.
We consider $\sigma^2\rightarrow K^2+\sigma^2$ to describe an unaccounted noise. We draw $K$ from the uniform distribution $U(0^\circ,10^\circ)$.
We also consider the possibility $\sigma^2\rightarrow K^2\sigma^2$ with $K$ drawn from $U(1,10)$. Both of these give similar fits so we present the analysis for 
$K^2+\sigma^2$ only. 
Note that the errors $\sigma$ in the phase shift data are estimated to be around $\sim 1^\circ\mbox{--}3^\circ$~\cite{Boykin:1972}.  Alternatively, one could also explore the possibility of an uncertainty in the overall normalization of the phase shift measurements. As the $s$ and $p$ wave phase shifts are from the same measurement and analysis, we did not consider the possibility of introducing separate measurement errors of the different elastic scattering channels.  
We use the following uniform prior distributions for the parameters and couplings in the EFT expressions:
%------------ Equation 10 --------------
\begin{align}
&a_0\sim U(1\ \mathrm{fm}, 70\ \mathrm{fm})\, , \nonumber\\
 &r_0\sim U(-5\ \mathrm{fm}, 5\ \mathrm{fm})\, , \nonumber\\
&s_0\sim U(-30\ \mathrm{fm^3}, 30\ \mathrm{fm^3})\, ,\nonumber\\
&\rho_1^{(+)}\sim U(-300\ \mathrm{MeV}, -48\ \mathrm{MeV})\, , \nonumber\\
&\rho_1^{(-)}\sim U(-300\ \mathrm{MeV}, -33\ \mathrm{MeV})\, , \nonumber\\
&\sigma_1^{(\pm)}\sim U(-5\ \mathrm{fm}, 5\ \mathrm{fm})\, ,\nonumber\\
&L_1^{(\pm)}\sim U(-10, 10)\, ,\nonumber\\
&K\sim U(0^\circ,10^\circ)\, . 
\end{align}
Fits without phase shifts do not depend on $\sigma_1^{(\pm)}$ and $K$. Model B without phase shift data also does not depend on $s_0$. 
The range for each of the uniform distributions above was guided by EFT power counting estimates. The ranges were wide enough that about 95\% of the posterior distributions of the parameters and couplings are
not pressed against the boundaries. For certain parameters such as the $p$ wave effective ranges $\rho_1^{(\pm)}$ physical constraint that the ANCs be positive determines the upper bounds. All these assumptions are considered part of the background information in the proposition $H$ in the probability distributions.

%%%------------- Figure  3 --------------------------------
\begin{figure}[tbh]
\begin{center}
\includegraphics[width=0.47\textwidth,clip=true]{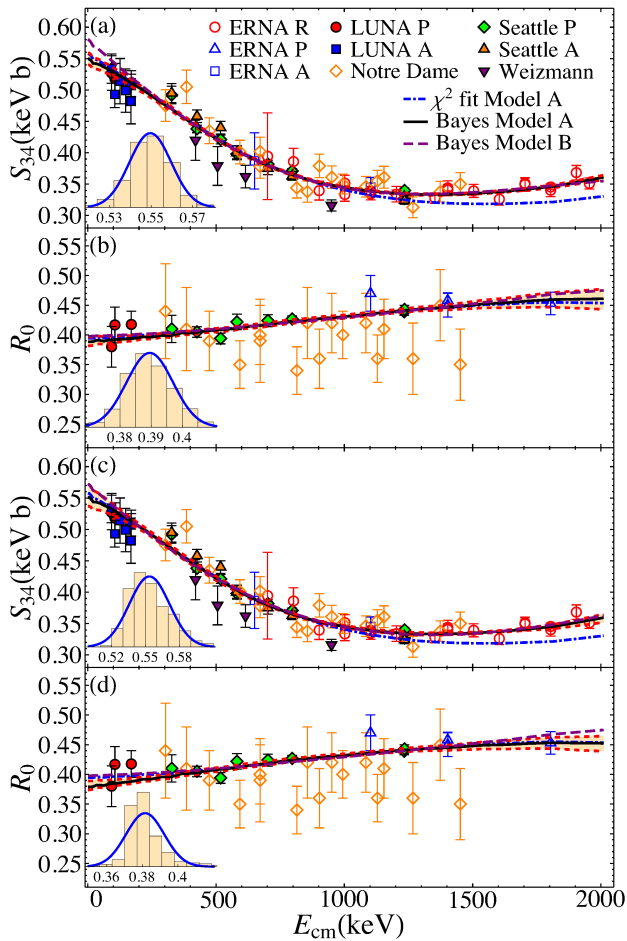}
\end{center}
\caption{\protect \HeAlpha: Model comparison for capture cross section. Notations the same as in Fig.~\ref{fig:Be7capture}. The fits included capture data in region II,   $E\lesssim 2000$ keV. }
\label{fig:Be7captureExt}
\end{figure}

%%%------------- Figure  4 --------------------------------
\begin{figure}[tbh]
\begin{center}
\includegraphics[width=0.47\textwidth,clip=true]{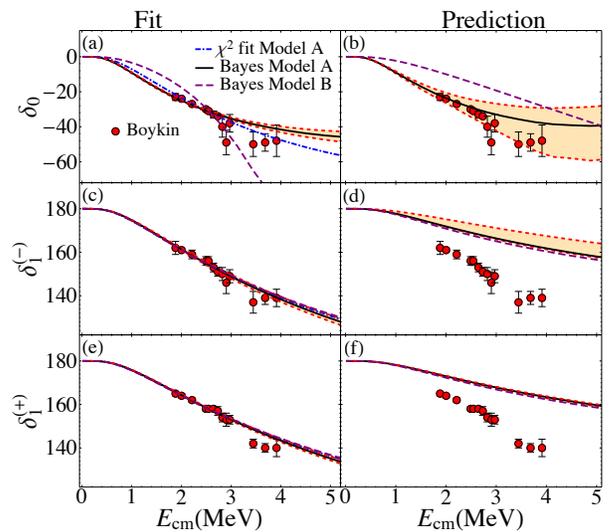}
\end{center}
\caption{\protect\HeAlpha: Model comparison for scattering phase shifts: Notations  
the same as in Fig.~\ref{fig:Be7PhaseShift} but with  capture data in region II $E\lesssim 2000$ keV. }
\label{fig:Be7PhaseShiftExt}
\end{figure}

In Figs.~\ref{fig:Be7capture} and \ref{fig:Be7PhaseShift}, we present fits in region I with and without phase shifts. In Fig.~\ref{fig:Be7capture}, the upper panels (a) and (b)  are fits with phase shift data, and the lower panels (c) and (d) are without. The panels on the left in Fig.~\ref{fig:Be7PhaseShift} are fits to phase shift data. The ones on the right are phase shift predictions from capture data. We include the $\chi^2$ fit of Model A to data in region I (with phase shift) for comparison~\cite{Higa:2016igc}. The parameter estimates are in Table~\ref{table:fitsBe7}.
The Bayesian curves are drawn from the posterior distributions of the parameters, not directly from the median values from the table. 
The posterior odd for the fit with phase shift is  $\ln[P(M_A|D,H)/P(M_B|D,H)]\approx2.6\pm0.8$ slightly favoring Model A, though the odds are similar if one accounts for systematic error. 
For the fit without phase shift data $\ln[P(M_A^\ast|D,H)/P(M_B^\ast|D,H)]\approx4.9\pm0.6$ is largely in favor of model A. We remind the reader that the asterisk $\ast$ indicates fits without phase shift data. From  Fig.~\ref{fig:Be7capture} panel (a), 
we see Model A (with phase shift) reproduces capture data in region II better than the other fits though it was fitted only to region I data.

Figs.~\ref{fig:Be7captureExt} and \ref{fig:Be7PhaseShiftExt} represent a similar analysis but fitted to capture data in region II. The parameter estimates are in Table~\ref{table:fitsBe7}. The posterior odds are $\ln[P(M_A|D,H)/P(M_B|D,H)]\approx31.6\pm0.9$ and $\ln[P(M_A^\ast|D,H)/P(M_B^\ast|D,H)]\approx0.2\pm0.8$. These fits indicate that Model B is not able to reproduce both phase shift and capture data simultaneously over a large region. The large $K$ value in Table ~\ref{table:fitsBe7} for Model B II corroborates this. Without constraint from phase shift, model A$^\ast$ II and model B$^\ast$ II are equally favored. 

%===================================================
\subsection{Capture reaction \texorpdfstring{\TrAlphaHead}{H3-Alpha}}

%%%------------- Figure  5 --------------------------------
\begin{figure}[tbh]
\begin{center}
\includegraphics[width=0.47\textwidth,clip=true]{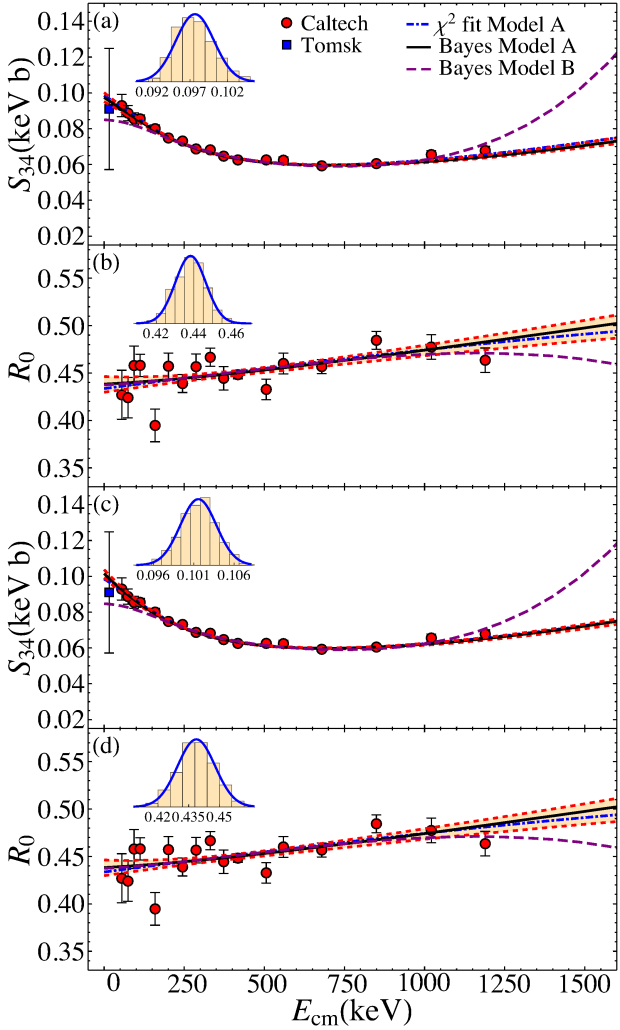}
\end{center}
\caption{\protect \TrAlpha: Model comparison for capture cross section. Solid (black) curve is the median result using Model A. The shaded band around it, bounded by the dashed (red) curves, represents $68\%$ of the posterior distribution. The long-dashed (purple) curve is for Model B. 
  The dot-dashed (blue) curve is for $\chi^2$ fit.
The inset shows Model A at $E_\star$.}
\label{fig:Li7capture}
\end{figure}

%%%------------- Figure  6--------------------------------
\begin{figure}[tbh]
\begin{center}
\includegraphics[width=0.47\textwidth,clip=true]{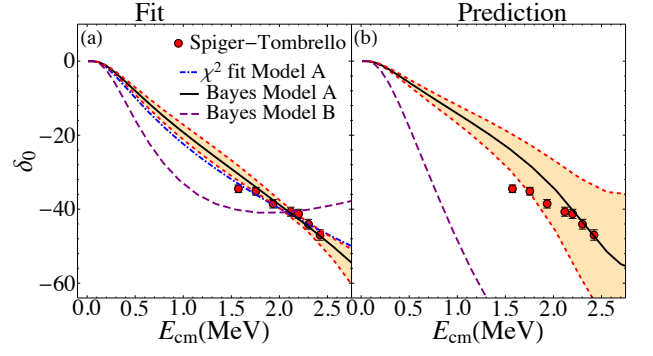}
\end{center}
\caption{\protect\TrAlpha: Model comparison for  scattering phase shifts. Notation for the curves 
the same as in Fig.~\ref{fig:Li7capture}. The panel on the left is fit to phase shift data whereas the panel on the right are predictions.}
\label{fig:Li7PhaseShift}
\end{figure}

In the \TrAlpha ~calculation we use $Z_\psi=1$ and mass $M_\psi=2809.43$ MeV for the incoming $^3$H. The charge and mass for the $\alpha$ is the same as before. The data for the capture cross section are from Caltech~\cite{Brune:1994} and Tomsk~\cite{Tomsk:2017}.
As in the case of $^7$Be, if we associate the binding momenta  for the ground and 1st excited of $^7$Li $\gamma_0\sim\gamma_1\sim80\mbox{--}90$ MeV with the scale $Q$, then for this system we expect the convergence to be slower compared to the previous $^3$He-$\alpha$ system which had smaller $\gamma_0$, $\gamma_1$. 

The role of the initial state interaction and the two-body current contributions are predicated on the size of the $s$-wave scattering length $a_0$ as before. However, in this system a larger $a_0\sim \Lambda^2/Q^3$ corresponds to only $\sim10\mbox{--}20\ \mathrm{fm}$. A smaller $a_0\sim \Lambda/Q^2\sim 1\mbox{--}5\ \mathrm{fm}$. The larger $a_0$ would make the initial state strong interaction and two-body current contributions LO, and the smaller $a_0$  would make these contributions NLO.  The power-counting arguments for \TrAlpha ~are similar to the \HeAlpha ~capture. The difference is that now $Q$ is numerically larger. 

The phase shift information of the initial $^3$H-$\alpha$ state is less well known than the $^3$He-$\alpha$ state. We digitized the $s$-wave phase shift $\delta_0$ for momentum range $p\sim70\mbox{--}90$ MeV from Fig. 15 in Ref.~\cite{Spiger:1967}. The $p$ waves appear very noisy and we decided not to use these in the fits. The authors estimated an error of about $5^\circ$ in the phase shift analysis for $^3$He-$\alpha$ but the errors in the $^3$H-$\alpha$ phase shifts are not listed. We find a mention of 3\% systematic error in $^3$H-$\alpha$ that we use. It is likely the error in this system is similar to $^3$He-$\alpha$. We add an additional noise as before and use $K^2+\sigma^2$ in the fits with $K\sim U(0^\circ,10^\circ)$. 
We use the following priors:
%-------------- Equation 11 --------------
\begin{align}
&a_0\sim U(1\ \mathrm{fm}, 70\ \mathrm{fm})\, , \nonumber\\
 &r_0\sim U(-5\ \mathrm{fm}, 5\ \mathrm{fm})\, , \nonumber\\
&s_0\sim U(-45\ \mathrm{fm^3}, 45\ \mathrm{fm^3})\, ,\nonumber\\
&\rho_1^{(+)}\sim U(-320\ \mathrm{MeV}, -126\ \mathrm{MeV})\, , \nonumber\\
&\rho_1^{(-)}\sim U(-320\ \mathrm{MeV}, -106\ \mathrm{MeV})\, , \nonumber\\
&L_1^{(\pm)}\sim U(-10, 10)\, ,\nonumber\\
&K\sim U(0^\circ,10^\circ)\, . 
\end{align}
The fits do not depend on the $p$-wave shape parameters $\sigma_1^{(\pm)}$ since we do not use the $p$-wave phase shifts. The upper limits on the effective ranges $\rho_1^{(\pm)}$ are constrained by requiring positive ANCs. We use 18 $S$-factor $S_{34}$ data, 17 branching ratio $R_0$ data and $7$ phase shift $\delta_0$ data points for the fits. 

Figs.~\ref{fig:Li7capture} and ~\ref{fig:Li7PhaseShift} shows the Bayesian fits. The power counting for Model A and Model B are the same as in \HeAlpha. We include a $\chi^2$ fit with phase shift using Model A for comparison. The upper panels (a) and (b) of 
Fig.~\ref{fig:Li7capture} are for fits with $s$-wave phase shift, and the lower panels (c) and (d) are without. Fig.~\ref{fig:Li7PhaseShift}, left panel are fits to phase shift whereas the right panel are the predictions from fits to capture data. The posterior odds overwhelmingly favor Model A with $\ln[P(M_A|D,H)/P(M_B|D,H)]\approx19.7\pm0.7$, $\ln[P(M_A^\ast|D,H)/P(M_B^\ast|D,H)]\approx18.4\pm0.6$. The parameter estimations are in Table~\ref{table:fitsLi7}. The fits to phase shift suggests errors of about 
$5^\circ$ in $\delta_0$.

%===================================================
\section{Discussion and Summary}
\label{sec:summary}

We performed several analysis of the EFT cross sections for \HeAlpha ~ and 
\TrAlpha ~to draw Bayesian inference. Two EFT power counting proposals were compared.  Typically, LO cross sections in EFT do not depend on two-body currents. This allows one to perform capture calculations once a few LO couplings are determined from elastic scattering data if they are known. For the two capture reactions considered here, elastic scattering data is not well measured. Available data on elastic scattering in $^3$He-$\alpha$~\cite{Boykin:1972,Spiger:1967} and $^3$H-$\alpha$~\cite{Spiger:1967} systems suggest the $s$-wave scattering lengths $a_0$s are fine tuned to be large. This makes initial state interactions in the capture reactions dominant, and as a consequence the two-body currents could contribute at LO~\cite{Higa:2016igc}. We called this EFT power counting Model A. Alternatively, one could have a power counting where two-body currents appear at higher order in the perturbation. This would require a smaller $s$-wave scattering length. We call this power counting Model B. If we ignore the poorly known elastic scattering data, then this second power counting can be used to describe the capture data. The relatively smaller initial state interaction contributions in Model B can be compensated by a larger wave function normalization constant which requires only small variation in the respective $p$-wave effective ranges as described earlier. We discuss the results of the analysis for \HeAlpha ~ and 
\TrAlpha ~separately below.

\subsection{\texorpdfstring{\HeAlphaHead}{He3-Alpha}}

We present the 8 different fits used to draw Bayesian inference for this reaction. First we start with the fits in the smaller capture energy region I ($E\lesssim 1000$ keV). The posterior odds favored Model A both with and without phase shift data in the fits. However, if we also look at the overall trend then we see the data for $S$-factor $S_{34}$ rises upward from around $E\sim 1500$ keV. The Model A fit with phase shift, we call Model A I, best describes the capture data over the energy range $E\lesssim 2000$ keV. We note that the Model B fits in this region are not self-consistent in that they suggest an $a_0$ value larger than the power counting, table~\ref{table:fitsBe7}.  

In the fits to capture energy region II ($E\lesssim 2000$ keV), Model A with phase shift, we call Model A II,  is favored over Model B by the posterior odd. For the fits without phase shifts, both Model A$^\ast$ II and Model B$^\ast$ II are equally favored by the posterior odd ratio. The asterisk $\ast$ indicates fits without phase shift data.

%------------- Table 1 -----------------------
\begin{table}[htb]
\centering
\begin{ruledtabular}
  \begin{tabular}{lll}
       \ Fit & $S_{34}(E_\star)$ (keV\ b) & $S'_{34}(E_\star)$ ($10^{-4}$\ b)\\ \hline
       \begin{filecontents*}{table_S34_Be7_dwave.csv}
Row,Sfactor,Up,Down,EFT,Sp,Spp,Spm,EFTB
$\chi^2$,0.558,0.008,0,0.056,-2.71,0.20,0,0.27
Model A I,0.541,0.012,0.014,0.054,-1.34,0.64,0.59,0.13
Model A II,0.550,0.009,0.010,0.055,-2.00,0.36,0.35,0.20
Model A$^\ast$ II,0.551,0.021,0.014,0.055,-1.86,0.72,1.69,0.19
Model B$^\ast$ II,0.573,0.007,0.007,0.017,-3.72,0.11,0.10,0.11
\end{filecontents*}
\csvreader[head to column names, late after line=\\]{table_S34_Be7_dwave.csv}{}
{\ \Row 
&\ifthenelse{\equal{\Down}{0}}{\Sfactor\ $\pm$ \Up $\pm \EFT$}{\Sfactor$^{+\Up}_{-\Down}\ \pm \EFT$} &\ifthenelse{\equal{\Spm}{0}}{$\Sp\pm\Spp\pm\EFTB$}{$\Sp^{+\Spp}_{-\Spm}\pm \EFTB$ }
}
     \end{tabular}
\end{ruledtabular}
\caption{\HeAlpha: $S_{34}$ and $S_{34}'$ at threshold (defined as $E_\star=60\times 10^{-3}$ keV). The second set of errors are estimated from the EFT perturbation as detailed in the text. 
}  
\label{table:S34Be7}
\end{table}  

%------------- Table 2 -----------------------
\begin{table}[htb]
\centering
\begin{ruledtabular}
  \begin{tabular}{ll}
       \ Fit & $R_0$ \\ \hline
       \begin{filecontents*}{table_Ratio_Be7_dwave.csv}
Row,Ratio,Up,Down,EFT
$\chi^2$,0.395,0.014,0,0.039
Model A I,0.387,0.018,0.015,0.039
Model A II,0.389,0.008,0.007,0.039
Model A$^\ast$ II,0.379,0.010,0.005,0.038
Model B$^\ast$ II,0.401,0.005,0.005,0.012
\end{filecontents*}
\csvreader[head to column names, late after line=\\]{table_Ratio_Be7_dwave.csv}{}
{\ \Row 
&\ifthenelse{\equal{\Down}{0}}{\Ratio\ $\pm$ \Up $\pm \EFT$}{\Ratio$^{+\Up}_{-\Down}\ \pm \EFT$}
}
     \end{tabular}
\end{ruledtabular}
\caption{\HeAlpha: Branching ratio $R_0$ at threshold (defined as $E_\star=60\times 10^{-3}$ keV). 
The second set of errors are estimated from the EFT perturbation as detailed in the text. 
}  
\label{table:RatioBe7}
\end{table}  

Tables~\ref{table:S34Be7} and ~\ref{table:RatioBe7} has the $S$-factors $S_{34}$ and branching ratios $R_0$ for the 4 fits described above. We also include the $\chi^2$ fit of Model A for comparison~\cite{Higa:2016igc}. We include the derivative $S'_{34}$ as well. All the numbers were evaluated at $E_\star=60\times 10^{-3}$ keV. We include the estimated EFT errors. The NLO Model A results have a 10\% error, and the NNLO Model B results have a 3\% error. The different EFT error estimates has to do with the distinction between ``accuracy and precision". The error estimates from higher order corrections represent precision, and different power countings have different accuracy and precision.

%%%------------- Figure  7 --------------------------------
\begin{figure}[tbh]
\begin{center}
\includegraphics[width=0.47\textwidth,clip=true]{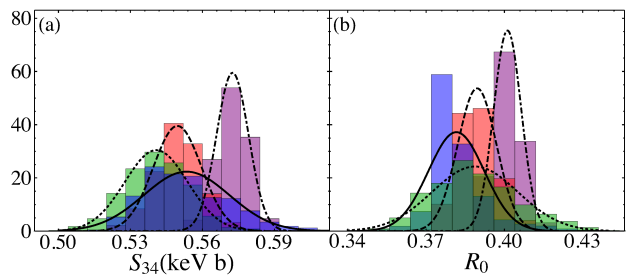} 
\end{center}
\caption{\protect\HeAlpha: $S$-factor $S_{34}$ and branching ratio $R_0$ at threshold ($E_\star=60\times 10^{-3}$ keV). The posterior distributions for fits Model A I, Model A II, Model A$^\ast$ II and Model B$^\ast$ II are presented as green, red, blue and purple colored histograms, respectively. The corresponding Gaussian probability distributions are given by the dotted, dashed, solid and dot-dashed curves, respectively.}
\label{fig:Be7Histogram}
\end{figure}

Fig.~\ref{fig:Be7Histogram} shows the posterior distributions for the 4 $S$-factors and branching ratios from 
Tables~\ref{table:S34Be7} and ~\ref{table:RatioBe7}. The symmetric distributions can be described with a Gaussian form shown by the various smooth curves unlike the skewed distributions as expected.  The spread in some of the quantities is related to the uncertainty in the parameter estimates, especially the $p$-wave effective ranges $\rho_1^{(\pm)}$ for Model A fits, see Table~\ref{table:fitsBe7}. Large magnitude  $|\rho_1^{(\pm)}|$ makes the wave function normalization constant smaller which can be compensated by a larger two-body current $L_1^{(\pm)}$ as the parameter estimates indicate.  The planned TRIUMF $^3$He-$\alpha$ elastic scattering experiments and phase shift analysis at low energies $E\gtrsim 500$ keV would be able to shed some light on this~\cite{Davids:TRIUMF}, and help establish the appropriate EFT power counting.   

The EFT $S$-factor at threshold  can be compared to other recent calculations such as -- $0.593$ keV b from 
FMD~\cite{Neff:2010nm}, $0.59$ keV b from  
NCSM~\cite{Dohet-Eraly:2015ooa}; and evaluations such as --
$[0.580\pm0.043 (\mathrm{stat.})\pm0.054 (\mathrm{sys.})]$ keV b from Cyburt-Davids~\cite{Cyburt:2008up}, 
$(0.57\pm0.04)$ keV b from 
ERNA~\cite{ERNA}, $(0.567\pm0.018\pm0.004)$ keV b 
from LUNA~\cite{LUNA}, $(0.554\pm0.020)$ keV b from Notre 
Dame~\cite{NotreDame}, $(0.595\pm0.018)$ keV b from 
Seattle~\cite{Seattle}, and  $(0.53\pm0.02\pm0.01)$ keV b from 
Weizmann~\cite{Weizmann}. We also compare to a recent EFT work using Bayesian inference~\cite{Zhang:2018qhm} that finds at threshold $S_{34}(0)=0.578^{+0.015}_{-0.016}$ keV b and $R_0(0)=0.406^{+0.013}_{-0.011}$. Ref.~\cite{Zhang:2018qhm} only used capture data to draw their inferences. Looking at tables ~\ref{table:S34Be7} and ~\ref{table:RatioBe7}, it would seem the results of Ref.~\cite{Zhang:2018qhm} are more aligned with Model B$^\ast$ II, though the exact power counting used the article is not clear.  
The ``best" recommended value from the review in Ref.~\cite{Adelberger:2010qa} is: $S_{34}(0)= [0.56\pm0.02(\mathrm{expt.})\pm0.02(\mathrm{theory})]$ keV b. 

From the various fits, we recommend the following: Model A II if we want to include phase shift information, and either Model A$^\ast$ II or Model B$^\ast$ II if no phase shift information is used.

\subsection{\texorpdfstring{\TrAlphaHead}{H3-Alpha}}

%------------- Table 3 -----------------------
\begin{table}[htb]
\centering
\begin{ruledtabular}
  \begin{tabular}{lll}
       \ Fit & $S_{34}(E_\star)$ (keV\ b) & $S'_{34}(E_\star)$ ($10^{-4}$\ b) \\ \hline
       \begin{filecontents*}{table_S34_Li7_dwave.csv}
Row,Sfactor,Up,Down,EFT,Sp,Spp,Spm,EFTB
$\chi^2$,0.098,0.003,0,0.016,-1.13,0.24,0,0.18
Model A,0.097,0.003,0.002,0.016,-1.04,0.12,0.20,0.17
Model A$^\ast$,0.102,0.002,0.002,0.016,-1.81,0.16,0.15,0.29
\end{filecontents*}
\csvreader[head to column names, late after line=\\]{table_S34_Li7_dwave.csv}{}
{\ \Row 
&\ifthenelse{\equal{\Down}{0}}{\Sfactor\ $\pm$ \Up $\pm \EFT$}{\Sfactor$^{+\Up}_{-\Down}\ \pm \EFT$} &\ifthenelse{\equal{\Spm}{0}}{$\Sp\pm\Spp\pm\EFTB$}{$\Sp^{+\Spp}_{-\Spm}\pm \EFTB$ }
}
     \end{tabular}
\end{ruledtabular}
\caption{\TrAlpha: $S_{34}$ and $S_{34}'$ at threshold (defined as $E_\star=60\times 10^{-3}$ keV). The second set of errors are estimated from the EFT perturbation as detailed in the text. 
}  
\label{table:S34Li7}
\end{table}  

\begin{figure}[tbh]
\begin{center}
\includegraphics[width=0.47\textwidth,clip=true]{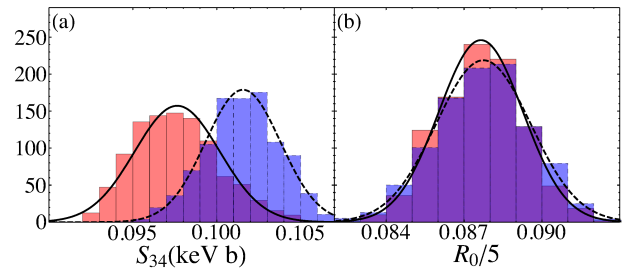} 
\end{center}
\caption{\protect\TrAlpha: $S$-factor $S_{34}$ and branching ratio $R_0$ at threshold ($E_\star=60\times 10^{-3}$ keV). The posterior distributions for fits Model A and Model A$^\ast$
are presented as red and  blue colored histograms, respectively. The corresponding Gaussian probability distributions are given by the solid and dashed curves, respectively.}
\label{fig:Li7Histogram}
\end{figure}

To draw our Bayesian inferences, we performed 4 different fits in this system. Here the EFT power counting of Model A (with phase shift data) and Model A$^\ast$ (without phase shift data) are favored. Moreover, the Model B and Model B$^\ast$ fits are not consistent with the power counting estimate for the $a_0$ values, table~\ref{table:fitsLi7}. The $S$-factors and branching ratios are in Tables~\ref{table:S34Li7} and \ref{table:RatioLi7}. The corresponding posterior distributions, and the associated Gaussian forms are shown in Fig.~\ref{fig:Li7Histogram}. The fit without phase shift data gives a larger $S_{34}$ at threshold.  We included a 16\% EFT error estimate from NNLO corrections. In this channel the binding energies are larger, resulting in a larger perturbation error compared to \HeAlpha.

The fits, especially without the phase shift data, give a central $s$-wave shape parameter value $s_0$ that is larger than ideally expected from the power counting. Power counting consistency would bias one towards Model A (with phase shift) over Model A$^\ast$ (without phase shift). 

The EFT $S$-factor calculation can be compared to recent theoretical results -- $0.12$ keV b from 
FMD~\cite{Neff:2010nm}, $0.13$ keV b from  
NCSM~\cite{Dohet-Eraly:2015ooa}; and experimental evaluation -- $[0.1067\pm0.0004 (\mathrm{stat.})\pm0.0060 (\mathrm{sys.})]$ keV b from Caltech~\cite{Brune:1994}. 
%------------- Table 4 -----------------------
\begin{table}[htb]
\centering
\begin{ruledtabular}
  \begin{tabular}{ll}
       \ Fit & $R_0$ \\ \hline
       \begin{filecontents*}{table_Ratio_Li7_dwave.csv}
Row,Ratio,Up,Down,EFT
$\chi^2$,0.434,0.006,0,0.069
Model A,0.438,0.008,0.009,0.070
Model A$^\ast$,0.439,0.008,0.009,0.070
\end{filecontents*}
\csvreader[head to column names, late after line=\\]{table_Ratio_Li7_dwave.csv}{}
{\ \Row 
&\ifthenelse{\equal{\Down}{0}}{\Ratio\ $\pm$ \Up $\pm \EFT$}{\Ratio$^{+\Up}_{-\Down}\ \pm \EFT$}
}
     \end{tabular}
\end{ruledtabular}
\caption{\TrAlpha: Branching ratio $R_0$ at threshold (defined as $E_\star=60\times 10^{-3}$ keV). 
The second set of errors are estimated from the EFT perturbation as detailed in the text.
}  
\label{table:RatioLi7}
\end{table}

%%%------------- Figure  8 --------------------------------

We leave for future a more careful study of the systematic errors associated with the estimation of the parameters and evidences using the various Nested Sampling methods. In our evaluations, for example, the 
MultiNest~\cite{Feroz:2009} and Diffusive Nested Sampling~\cite{Brewer:2011} algorithms gave similar results. The EFT parameters 
(tables ~\ref{table:fitsBe7} and ~\ref{table:fitsLi7}) agreed within the error bars. Similarly, the capture cross sections and branching ratios (tables ~\ref{table:S34Be7}, 
~\ref{table:RatioBe7}, ~\ref{table:S34Li7}, ~\ref{table:RatioLi7}) also agreed within the error bars from the fits.

%=============================x======================
\begin{acknowledgments}
The authors thank Barry Davids, Renato Higa, Daniel R. Phillips, Prakash Patil and Xilin Zhang for many valuable discussions. 
This work was supported in part by 
U.S. NSF grants PHY-1615092 and PHY-1913620.
The figures for this article have been created using SciDraw~\cite{SciDraw}. 
\end{acknowledgments}

\appendix

%====================== Appendix A ======================
\section{\texorpdfstring{$\bm{s}$ and $\bm{d}$ wave capture}{s and d wave capture}}
\label{sec:sd_wave}
The capture from initial $s$-wave state is given by:
%-----------  Equation --------------
\begin{align}\label{eq:swaveA}
 \frac{|\mathcal{A}(p)|}{C_0(\eta_p)}=\left|X(p)
    -\frac{2\pi }{\mu^2}
    \frac{B(p) +\mu J_0(p) +\mu^2 k_0 L^{(\zeta)}_{E1}}
    {[C_0(\eta_p)]^2 p (\cot\delta_0-i)}\right|\, , 
\end{align}
where $k_0$ is the photon energy and $L^{(\zeta)}_{E1}$ for $\zeta={}^2P_{3/2}$ and 
$\zeta={}^2P_{1/2}$ are the 2 two-body currents. We modified some definitions compared to 
Ref.~\cite{Higa:2016igc} but they are equivalent expressions. 

The contribution without initial state strong interaction is given by
%-----------  Equation --------------
\begin{multline}
X(p)=1+
\frac{2\gamma}{3}\frac{\Gamma(2+k_C/\gamma)}{C_0(\eta_p)}\\
\times\int_0^\infty dr\ r 
W_{-\frac{k_C}{\gamma},\frac{3}{2}}(2\gamma r)\partial_r
\left[ \frac{F_0(\eta_p)}{p r} \right],
\end{multline}
where $\gamma$ is the binding momentum of the ground or first excited state as appropriate. The momentum $k_C=\alpha_e Z_\psi Z_\phi\mu$ is the inverse of the Bohr radius. 
The Coulomb  expressions are:
%-----------  Equation --------------
\begin{align}
F_l(\eta_p,\rho)&=C_l(\eta_p)2^{-l-1}(-i)^{l+1}M_{i\eta_p, l+1/2}(2i\rho)\, ,\nonumber\\
C_l(\eta_p)&=\frac{2^l e^{-\pi\eta_p/2}|\Gamma(l+1+i\eta_p)|}
{\Gamma(2l+2)}\, ,
\end{align}
with conventionally defined  Whittaker functions  
$M_{k,\mu}(z)$ and $W_{k,\mu}(z)$. 
$F_l(\eta_p,\rho)$ is the regular Coulomb wave function.

The initial $s$-wave scattering in Eq.~(\ref{eq:swaveA}) is contained in the Coulomb subtracted phase shift $\delta_0$ parameterized by the modified effective range expansion
%-----------  Equation --------------
\begin{align}\label{eq:swaveERE}
[C_0(\eta_p)]^2 p (\cot\delta_0-i) =  -\frac{1}{a_0}+\frac{r_0}{2} p^2+\frac{s_0}{4}  p^4+\cdots \nonumber \\
-2k_C H(\eta_p)\, ,\nonumber\\
H(x)= \psi(ix)+\frac{1}{2ix}-\ln(ix)\, .
\end{align}
$\psi(x)$ is the digamma function, and the $\cdots$ represents terms with higher powers in $p^2$. The combination $B(p)+\mu J_0(p)$ can be evaluated as
%-----------  Equation --------------
\begin{multline}
B(p)+\mu J_0(p) =\frac{\mu^2}{3\pi}\frac{ip^3-\gamma^3}{p^2+\gamma^2}+k_C C(p)+\Delta B(p)\\
-\frac{k_C\mu^2}{2\pi}\left[2 H(\eta_p)+2\gamma_E -\frac{5}{3}+ \ln 4\pi
\right]\, .
\end{multline}
The function $C(p)$ is given by the double integral
%-----------  Equation --------------
\begin{multline}
     C(p)=\frac{\mu^2}{6\pi^2(p^2+\gamma^2)}\int_0^1dx\int_0^1dy
  \frac{1}{\sqrt{x(1-x)}\sqrt{1-y}}\\
  \times \left( xp^2\ln\left[\frac{\pi}{4k_C^2}
    (-yp^2+(1-y)\gamma^2/x-i\delta) \right] \right.\\
  \left. +p^2 \ln\left[\frac{\pi}{4k_C^2}
    (-yp^2-(1-y)p^2/x-i\delta) \right]  \right.\\
  \left. +x\gamma^2\ln\left[\frac{\pi}{4k_C^2}
    (y\gamma^2+(1-y)\gamma^2/x-i\delta) \right]  \right.\\
  \left. +\gamma^2\ln\left[\frac{\pi}{4k_C^2}(y\gamma^2-(1-y)
    p^2/x-i\delta) \right] 
  \right)\, ,
\end{multline}
which is reduced to a single integral before evaluating numerically.  The function $\Delta B(p)$ is obtained from 
%-----------  Equation --------------
\begin{multline}
  B_{ab}(p)=-3\int d^3 r
  \left[\frac{G_C^{(1)}(-B;r',r)}{r'}\right]\Big|_{r'=0}\\
\times   \frac{\partial G_C^{(+)}(E;\bm{r},0)}{\partial r_a}\frac{r_b}{r}
 \,  ,\\
  \left[\frac{G_C^{(1)}(-B;r',r)}{r'}\right]\Big|_{r'=0}=-\frac{\mu\gamma}{6\pi r}
  \Gamma(2+k_C/\gamma)W_{-\frac{k_C}{\gamma},\frac{3}{2}}(2\gamma r)\, , \\
  G_C^{(+)}(E;\bm{r},0)=-\frac{\mu}{2\pi r}\Gamma(1+i\eta_p)
  W_{-i\eta_p,\frac{1}{2}}(-i2 p r)\, .    
\end{multline}
The integral $B_{ab}(p)$ is divergent at $r=0$. However, when combined with the contribution from $J(p)\delta_{ab}$ it is finite. Thus we make the substitution $(r_b/r)[\partial/\partial r_a]=(r_ar_b/r^2)[\partial/\partial r]
\rightarrow (\delta_{ab}/3)[\partial/\partial r]$ in the integral and accordingly 
$B_{ab}(p)\equiv B(p)\delta_{ab}$. The finite piece $\Delta B(p)$ is obtained numerically from $B(p)$ after subtracting the  zero and single photon contributions i.e. removing terms up to order $\alpha_e^2$~\cite{Higa:2016igc}.

The capture from initial $d$-wave states is given by the amplitude~\cite{Higa:2016igc}: 
%-----------  Equation --------------
\begin{multline}
\label{eq:dwaveA}
Y(p)=  \frac{2\gamma}{3}\Gamma(2+k_C/\gamma)\int_0^\infty dr\  r W_{-k_C/\gamma,3/2}(2\gamma r) \\
\times \left(\frac{\partial}{\partial r} +\frac{3}{r}\right) \frac{F_2(\eta_p, r p)}{r p}.
\end{multline}

%====================== Appendix B ======================
\section{Wave function renormalization}
\label{sec:Zphi}
The wave function renormalization constant is calculated as~\cite{Higa:2016igc}:
%-----------  Equation --------------
\begin{multline}\label{eq:Zphi}
    -\frac{2\pi}{h^{(\zeta)2}{\cal Z}^{(\zeta)}}
=\frac{1}{p}\frac{\partial}{\partial p} 9 [C_1(\eta_p)]^2 p^3(\cot\delta_1-i)\\
=\rho_1^{(\zeta)}-4k_C\,H\left(-i\frac{k_C}{\gamma}\right)
-\frac{2k_C^2}{\gamma^3}(k_C^2-\gamma^2)\\
\left[\psi'\left(\frac{k_C}{\gamma}\right)
-\frac{\gamma^2}{2k_C^2}-\frac{\gamma}{k_C}\right]\, ,
\end{multline}
where we used the relation 
\begin{multline}
9 [C_1(\eta_p)]^2 p^3(\cot\delta_1-i)= 
 2 k_C(k_C^2-\gamma^2)H(-i\eta_\gamma)\\+\frac{1}{2}\rho_1(p^2+\gamma^2)
+\frac{1}{4}\sigma_1(p^2+\gamma^2)^2+\cdots \\
 -2 k_C(k_C^2+p^2) H(\eta_p). 
\end{multline}

%====================== Appendix C ======================
\section{Parameter Estimates}
\label{sec:Fits}

We list the parameter estimates for both \HeAlpha ~and \TrAlpha ~below. For the Bayesian fits we show the median and the bounds containing 68\% of the posterior distributions.  Typically, a very asymmetric bound indicates either a skewed or sometimes a bi-modal distribution.
%------------- Table 5 ---------------------
%\begin{widetext}
%\begin{center}
\begin{table*}[tbh]
\centering
\begin{adjustbox}{width=0.97\textwidth,center}
%\begin{ruledtabular}
\begin{tabular}{|c|c|c|c|c|c|c|c|c|c|c|}\hline
Fits & $a_0$ (fm) & $r_0$ (fm) & $s_0$ (fm$^3$) &  $\rho_1^{(+)}$ (MeV) & $\sigma_1^{(+)}$ (fm)& $\rho_1^{(-)}$ (MeV) & $\sigma_1^{(-)}$  (fm) & $L_1^{(+)}$  & $L_1^{(-)}$ & $K$
\begin{filecontents*}{table_Be7_parameters.csv}
Fits,a,da,amid,ap,am,r,dr,rmid,rp,rm,sz,dsz,szmid,szp,szm,ra,dra,ramid,rap,ram,sa,dsa,samid,sap,sam,rb,drb,rbmid,rbp,rbm,sb,dsb,sbmid,sbp,sbm,La,dLa,Lamid,Lap,Lam,Lb,dLb,Lbmid,Lbp,Lbm,K,dK,Kmid,Kp,Km
$\chi^2$,22,3,0,0,0,1.2,0.1,0,0,0,-0.9,0.7,0,0,0,-55.4,0.5,0,0,0,1.59,0.03,0,0,0,-41.9,0.7,0,0,0,1.74,0.05,0,0,0,0.78,0.06,0,0,0,0.83,0.08,0,0,0,0,0,0,0,0
Model A I,48,2,48,2,2,1,0.09,1,0.09,0.1,-1.8,0.9,-1.8,1,0.9,-73,7,-72,5,8,2.1,0.2,2.1,0.2,0.2,-50,5,-49,3,6,2,0.2,2,0.2,0.1,1.4,0.1,1.4,0.2,0.1,1.3,0.2,1.2,0.2,0.1,0.4,0.3,0.3,0.4,0.2
Model B I,38,2,38,3,2,1.1,0.1,1.1,0.1,0.1,-2,1,-2,1,1,-61.6,0.5,-61.6,0.6,0.6,1.77,0.03,1.77,0.03,0.04,-50,5,-48,1,7,2,0.2,2,0.2,0.08,1.13,0.03,1.13,0.03,0.02,1.3,0.2,1.2,0.3,0.08,0.3,0.2,0.3,0.3,0.2
Model A$^\ast$ I,21,6,20,8,5,-0.2,0.6,-0.1,0.5,0.7,-16,6,-16,6,8,-90,20,-89,9,20,0,0,0,0,0,-130,50,-130,50,70,0,0,0,0,0,3.2,1,3,1,0.9,6,3,7,2,3,0,0,0,0,0
Model B$^\ast$ I,35,7,37,3,10,0.8,0.9,1.1,0.1,0.9,0,0,0,0,0,-61.3,0.9,-61.4,1,0.8,0,0,0,0,0,-60,30,-47,2,6,0,0,0,0,0,1.15,0.06,1.14,0.09,0.04,2,2,1.2,0.2,0.1,0,0,0,0,0
Model A II,40,5,40,5,6,1.1,0.1,1.09,0.09,0.1,-2.1,0.7,-2.2,0.8,0.8,-59,1,-59,1,2,1.69,0.05,1.69,0.05,0.06,-45,2,-45,2,2,1.84,0.07,1.84,0.08,0.08,1.02,0.05,1.02,0.06,0.06,1.07,0.08,1.07,0.08,0.09,0.3,0.2,0.3,0.3,0.2
Model B II,7.3,0.6,7.3,0.7,0.7,1.31,0.02,1.31,0.02,0.02,6,1,6,1,1,-53.5,0.1,-53.5,0.1,0.1,1.53,0.06,1.53,0.05,0.06,-40.1,0.2,-40.1,0.2,0.2,1.68,0.06,1.67,0.06,0.06,-0.1,0.1,-0.04,0.08,0.1,-0.02,0.09,-0.01,0.09,0.1,2.3,0.5,2.2,0.6,0.5
Model A$^\ast$ II,48,7,46,10,4,0.9,0.2,1,0.1,0.3,-2,3,-3,5,2,-62,6,-62,5,4,0,0,0,0,0,-70,40,-51,4,70,0,0,0,0,0,1.1,0.2,1.1,0.1,0.2,1.7,0.9,1.3,2,0.2,0,0,0,0,0
Model B$^\ast$ II,5,1,5,1,2,1.2,0.1,1.24,0.04,0.2,0,0,0,0,0,-53.5,0.1,-53.5,0.1,0.2,0,0,0,0,0,-40.2,0.2,-40.2,0.2,0.2,0,0,0,0,0,-0.8,0.8,-0.5,0.2,1,-0.7,0.8,-0.4,0.2,0.9,0,0,0,0,0
\end{filecontents*}
\csvreader[head to column names]{table_Be7_parameters.csv}{}
{ \\\hline\Fits &\ifthenelse{\equal{\amid}{0}}{$\a\pm\da$}{$\amid^{+\ap}_{-\am}$}  
& \ifthenelse{\equal{\rmid}{0}}{$\r\pm\dr$}{$\rmid^{+\rp}_{-\rm}$} 
 &\ifthenelse{\equal{\szmid}{0}}{\ifthenelse{\equal{\sz}{0}}{--}{$\sz\pm\dsz$}}{$\szmid^{+\szp}_{-\szm}$} 
& \ifthenelse{\equal{\ramid}{0}}{$\ra\pm\dra$}{$\ramid^{+\rap}_{-\ram}$}
& \ifthenelse{\equal{\samid}{0}}{\ifthenelse{\equal{\sa}{0}}{--}{$\sa\pm\dsa$}}{$\samid^{+\sap}_{-\sam}$} 
&\ifthenelse{\equal{\rbmid}{0}}{$\rb\pm\drb$}{$\rbmid^{+\rbp}_{-\rbm}$}
& \ifthenelse{\equal{\sbmid}{0}}{\ifthenelse{\equal{\sb}{0}}{--}{$\sb\pm\dsb$}}{$\sbmid^{+\sbp}_{-\sbm}$} 
& \ifthenelse{\equal{\Lamid}{0}}{$\La\pm\dLa$}{$\Lamid^{+\Lap}_{-\Lam}$} &\ifthenelse{\equal{\Lbmid}{0}}{$\Lb\pm\dLb$}{$\Lbmid^{+\Lbp}_{-\Lbm}$}
& \ifthenelse{\equal{\Kmid}{0}}{\ifthenelse{\equal{\K}{0}}{--}{$\K\pm\dK$}}{$\Kmid^{+\Kp}_{-\Km}$}}\\\hline
\end{tabular}
%\end{ruledtabular}
\end{adjustbox}
\caption{\HeAlpha: EFT parameters. We estimate the parameters from fits to capture data in region I ($E\lesssim 1000$ keV) and in region II ($E\lesssim 2000$ keV) as indicated. The fits that do not use elastic scattering phase shift data are indicated by the asterisk $\ast$ sign. We show the errors to one significant figure.}
 \label{table:fitsBe7}
\end{table*}  
%\end{center}
%\end{widetext}

%------------- Table 6 ---------------------
%\begin{widetext}
%\begin{center}
\begin{table*}[t]
\centering
\begin{adjustbox}{width=0.7\textwidth,center}
%\begin{ruledtabular}
\begin{tabular}{|c|c|c|c|c|c|c|c|c|}\hline
Fits & $a_0$ (fm) & $r_0$ (fm) & $s_0$ (fm$^3$) &  $\rho_1^{(+)}$ (MeV) 
& $\rho_1^{(-)}$ (MeV) 
& $L_1^{(+)}$ 
& $L_1^{(-)}$ & $K$
\begin{filecontents*}{table_Li7_parameters.csv}
Fits,a,da,amid,ap,am,r,dr,rmid,rp,rm,sz,dsz,szmid,szp,szm,ra,dra,ramid,rap,ram,rb,drb,rbmid,rbp,rbm,La,dLa,Lamid,Lap,Lam,Lb,dLb,Lbmid,Lbp,Lbm,K,dK,Kmid,Kp,Km
$\chi^2$,17,2,0,0,0,0.6,0.3,0,0,0,2,2,0,0,0,-149,3,0,0,0,-129,4,0,0,0,1.44,0.07,0,0,0,1.5,0.1,0,0,0,0,0,0,0,0
Model A,13,2,13,2,1,-0.1,0.7,0,0.5,0.8,13,8,11,10,7,-190,30,-190,30,30,-230,60,-230,80,60,2.3,0.6,2.2,0.7,0.6,4,1,4,1,2,2,2,2,2,1
Model B,21,2,21,2,2,2.7,0.1,2.7,0.1,0.1,-24,2,-24,3,2,-218,8,-219,9,8,-198,9,-200,10,10,2.15,0.02,2.14,0.01,0.01,2.4,0.1,2.4,0.1,0.1,6,2,6,2,2
Model A$^\ast$,10,2,10,2,2,-2.1,0.8,-2,0.8,0.9,35,7,36,6,10,-220,30,-220,30,30,-250,40,-240,50,50,3.4,0.7,3.3,0.9,0.8,5,1,5,2,1,0,0,0,0,0
Model B$^\ast$,20,1,20,1,1,2.73,0.09,2.7,0.1,0.1,0,0,0,0,0,-213,5,-213,6,4,-191,7,-191,8,6,2.14,0.02,2.14,0.01,0.01,2.37,0.09,2.37,0.1,0.09,0,0,0,0,0
\end{filecontents*}
\csvreader[head to column names]{table_Li7_parameters.csv}{}
{ \\\hline\Fits 
&\ifthenelse{\equal{\amid}{0}}{$\a\pm\da$}{$\amid^{+\ap}_{-\am}$}
& \ifthenelse{\equal{\rmid}{0}}{$\r\pm\dr$}{$\rmid^{+\rp}_{-\rm}$} 
&\ifthenelse{\equal{\szmid}{0}}{\ifthenelse{\equal{\sz}{0}}{--}{$\sz\pm\dsz$}}{$\szmid^{+\szp}_{-\szm}$} 
& \ifthenelse{\equal{\ramid}{0}}{$\ra\pm\dra$}{$\ramid^{+\rap}_{-\ram}$}
&\ifthenelse{\equal{\rbmid}{0}}{$\rb\pm\drb$}{$\rbmid^{+\rbp}_{-\rbm}$}
& \ifthenelse{\equal{\Lamid}{0}}{$\La\pm\dLa$}{$\Lamid^{+\Lap}_{-\Lam}$} &\ifthenelse{\equal{\Lbmid}{0}}{$\Lb\pm\dLb$}{$\Lbmid^{+\Lbp}_{-\Lbm}$}
& \ifthenelse{\equal{\Kmid}{0}}{\ifthenelse{\equal{\K}{0}}{--}{$\K\pm\dK$}}{$\Kmid^{+\Kp}_{-\Km}$}
}\\\hline
\end{tabular}
%\end{ruledtabular}
\end{adjustbox}
\caption{\TrAlpha: EFT parameters. The fits that do not use elastic scattering phase shift data are indicated by the asterisk $\ast$ sign. We show the errors to one significant figure.}
 \label{table:fitsLi7}
\end{table*}  
%\end{center}
%\end{widetext}

\bibliographystyle{apsrev4-1}
%\bibliography{Reference.bib}


\end{document}